\documentclass[english,twocolumn]{revtex4}
\usepackage[T1]{fontenc}
\usepackage[latin9]{inputenc}
\usepackage{float}
\usepackage{bm}
\usepackage{amsmath}
\usepackage{amssymb}
\usepackage{graphicx}
\usepackage{esint}

\makeatletter
\@ifundefined{textcolor}{}
{%
 \definecolor{BLACK}{gray}{0}
 \definecolor{WHITE}{gray}{1}
 \definecolor{RED}{rgb}{1,0,0}
 \definecolor{GREEN}{rgb}{0,1,0}
 \definecolor{BLUE}{rgb}{0,0,1}
 \definecolor{CYAN}{cmyk}{1,0,0,0}
 \definecolor{MAGENTA}{cmyk}{0,1,0,0}
 \definecolor{YELLOW}{cmyk}{0,0,1,0}
 }

\makeatother

\usepackage{babel}
\begin{document}

\title{Parametric Projection Operator Technique for Second Order Non-linear
Response}

\author{Jan Ol\v{s}ina and Tom\'{a}\v{s} Man\v{c}al}

\affiliation{Faculty of Mathematics and Physics, Charles University in Prague,
Ke Karlovu 5, CZ-121 16 Prague 2, Czech Republic}
\begin{abstract}
We demonstrate the application of the recently introduced parametric
projector operator technique to a calculation of the second order
non-linear optical response of a multilevel molecular system. We derive
a parametric quantum master equation (QME) for the time evolution
of the excited state of an excitonic system after excitation by the
first two pulses in the usual spectroscopic four-wave-mixing scheme.
This master equation differs from the usual QME by a correction term
which depends on the delay $\tau$ between the pulses. In the presence
of environmental degrees of freedom with finite bath correlation time
and in the presence of intramolecular vibrations we find distinct
dynamics of both the excite state populations and the electronic coherence
for different delays $\tau$.
\end{abstract}
\maketitle

\section{Introduction\label{sec:Introduction}}

Recent years have seen a rapid experimental development in multidimensional
coherent spectroscopy. Developed first in the nuclear magnetic resonance
\cite{ErnstBodenhausenWokaunBook} it has been brought to infra red
\cite{Hamm1998a,Asplund2000a,Cervetto2004a} and later into near infra-red
and optical domains \cite{Cowan2004a,Brixner2004b}. Since then it
yielded new insights into photo-induced dynamics of electronic excited
states of small molecules \cite{MIlota2009a}, polymers \cite{Collini2009a},
large photosynthetic aggregates \cite{Engel2007a,Ginsberg2009a} and
even solid state systems \cite{Stone2009a}. The two-dimensional (2D)
Fourier transformed spectrum completely characterizes the third order
non-linear response of a molecular ensemble in amplitude and phase
\cite{Mukamel2000a,Jonas2003a} providing thus, the maximal information
accessible in a three pulse experiment. In the three pulse degenerate
four wave mixing (FWM) experiment \cite{MukamelBook}, two independent
adjustable parameters -- the delays between the three pulses -- allow
experimentalists to address the time and delay dependent third order
non-linear response of the molecular system $S^{(3)}(t,T,\tau)$ via
complete characterization of the third order non-linear signal $E_{s}^{(3)}(t,T,\tau)$
(see Fig. \ref{fig:FWM-Experiment}). This information is conveniently
represented by a 2D plot of the Fourier transformed signal $E_{s}^{(3)}(\omega_{t},T,\omega_{\tau})$
which correlates the absorption frequency $\omega_{\tau}$ with the
frequency $\omega_{t}$ of stimulated emission, ground state bleaching
and excited state absorption contributions, separated from the absorption
event by an adjustable waiting time $T$. Femtosecond photo-induced
evolution of molecular systems is reflected in the amplitude, position
and shape of 2D spectral features as functions of the waiting time
revealing thus important information about the molecular structure,
intramolecular dynamics, as well as the interaction of molecular electronic
transitions with their environment \cite{Cho2005a,Pisliakov2006a,Kjellberg2005a}. 

By analyzing 2D spectra of photosynthetic Fenna-Mathews-Olson (FMO)
protein, Brixner et al. have first demonstrated the ability of the
new technique to improve our knowledge of energy transfer pathways
in photosynthetic proteins \cite{Brixner2005a,Cho2005a}. Later, signatures
of electronic quantum coherence in this system have been theoretically
predicted \cite{Pisliakov2006a} and experimentally verified \cite{Engel2007a}.
The same measurements led to a surprising discovery of the long life
time of these electronic coherence and to an identification of a host
of other effects which contribute to the wave-like nature of the energy
transfer in these systems. Recently, coherent oscillations in 2D spectra
of photosynthetic systems were demonstrated even for room temperature
\cite{Collini2009b,Panitchayangkoon2010a}. The impact of these new
finds has been felt well beyond the photosynthesis research community,
and the quantum properties of energy transfer in biological systems
have attracted researches from seemingly unrelated fields of quantum
computation and quantum information science. Questions about the relevance
of quantum effects in natural light harvesting have led researches
to study quantum entanglement \cite{Ishizaki2010a,Sarovar2010a,Caruso2010a},
various aspects of the environmental assistance in quantum transport
and optimality of transport processes \cite{Mohseni2008a,Castro2008a,Rebentrost2009a,Caruso2009a,Chin2010a}.
Meanwhile, measurement of electronic coherence have found utility
in determining structure-related properties of photosynthetic systems.
Thus, the electronic beating of 2D spectra have been recently used
to precisely determine electronic energy levels in congested spectra
of molecular aggregates \cite{Calhoun2009a}.

In order to yield the above discussed impressive results, optical
2D experiments have to be accompanied by a thorough analysis, which
requires detailed theoretical understanding of the molecular response
to exciting light. For the description of the most 2D experiments,
the third order semi-classical light-matter interaction response function
theory is well established. Response functions of model few-level
systems with pure dephasing (i.e. with no energy transfer between
the levels) can even be expressed analytically in terms of the so-called
\textit{\emph{energy gap correlation function}} (EGCF), using the
second order cumulant in Magnus expansion \cite{MukamelBook}. Some
examples of small chromophores in solution fall in this category \cite{Nemeth2008a,Nemeth2010a}
when investigated on time scales shorter then radiative life time.
For a Gaussian bath this analytical theory is exact, and thus knowing
the EGCF of the electronic transitions enables us to determine linear
(absorption) as well as non-linear spectra. 

However, the construction of exact response functions for realistic
energy transferring systems, such as Frenkel excitons in photosynthetic
aggregates \cite{ValkunasBook}, is no longer possible. Photosynthetic
complexes are relatively large, and the proper methods to simulate
finite timescale stochastic fluctuations at finite temperatures \cite{Tanimura2006a,Ishizaki2009c,Ishizaki2009b}
carry a substantial numerical cost. Practical calculations thus require
some type of reduced dynamics where only electronic degrees of freedom
(DOF) are treated explicitly. These approaches usually rely on a host
of approximations that seem to work well for most spectroscopic techniques
\cite{Novoderezhkin2003a,Novoderezhkin2005a,Grondelle2006a,Abramavicius2009a}
and in systems where the details of system-bath coupling are apparently
less important \cite{Zigmantas2006a}. With increasing details of
the excited state dynamics revealed by the 2D spectroscopy \cite{Read2007a,Rhee2009a,Abramavicius2010a}
and with increasing size of the studied systems, it becomes more and
more important to keep the numerical cost of simulations low, while
simultaneously account for experimentally observed quantum effects.
One of the possible research directions is to extend on existing reduced
density matrix (RDM) theories \cite{Zhang1998a,Palmieri2009a} or
relax certain approximations \cite{Olsina2010a}.

It was demonstrated that RDM master equations derived by projection
operator technique reproduce exactly the linear response \cite{Doll2008a}.
In the case of higher order response functions, however, the same
approach necessarily neglects bath correlations between the different
periods of photo-induced system evolution (i.e. between so-called
coherence time $\tau$ and the waiting time $T$). The failure to
account for this correlation leads sometimes to a complete loss of
the experimentally observed dynamics in simulated 2D spectra, such
as in the case of the vibrational modulation of electronic 2D spectra
\cite{Nemeth2008a}. Because vibrational modulation leads to effects
similar to those attributed to electronic coherence, developing methods
that can account for its effect reliably in complex systems is of
utmost importance. One possible approach to the problem is to derive
equations of motion for the response as a whole, and to take the previous
time evolution of the system into account explicitly \cite{Richter2010a}.
In this paper, we take a different route in which the correlation
effects are treated by a specific choice of the projection operator.
Once a projector is specified, it can be used for any method of treating
system bath interaction. We apply the previously suggested parametric
projection operator technique \cite{Mancal2010c} to a calculation
of the second order response of a quantum system. The second order
response operators can be used to determine the state of a molecular
system subject to excitation by a weak light with arbitrary properties,
and thus its importance goes beyond the semi-classical system light
interaction theory \cite{Mancal2010a}. Alternatively, calculation
of the second order response can be viewed as a first step towards
more involved calculation of the third order response functions that
is required for the modeling of the third order non-linear spectra.

The paper is organized as follows: in the next section we specify
the model Hamiltonian and the description of the system's interaction
with the light and its environment (the thermodynamic bath). In Section
\ref{sec:Multi-point-Correlation-Function} we discuss the third order
response functions of a multilevel systems, and we point out general
limitations of their evaluation by the reduced density matrix master
equations. In Section \ref{sec:Second-Order-Response} we write out
the density operator describing the excited state of a molecular system
in terms of the second order response operator, and we discuss the
advantages of usage of the parametric projection operator over the
standard projectors. The details of the parametric projection operator
and the corresponding master equation are introduced in Section \ref{sec:Reduced-density-matrix}.
Numerical results for the excited state dynamics are then discussed
in Section \ref{sec:Numerical-Results-and}. Some definitions and
calculation details can be found in the Appendices.

\section{Model System and Block Formalism\label{sec:Model-System-and}}

\begin{figure}[h]
\centering{}\includegraphics[width=1\columnwidth]{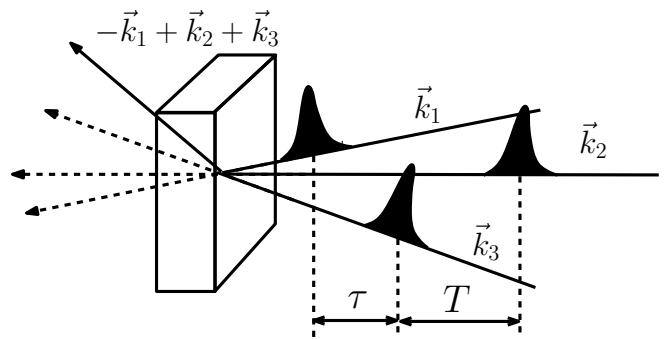} \caption{\label{fig:FWM-Experiment}A scheme of a typical FWM experiment. Three
laser pulses with delays $\tau$ and $T$ are incident on a sample.
A non-linear signal generated into a selected direction, $-\bm{k}_{1}+\bm{k}_{2}+\bm{k}_{3}$
here, is detected.}
\end{figure}

Our primary interest lies in photosynthetic aggregates of chlorophylls.
Frenkel exciton model is well established for the description of photosynthetic
systems and their spectroscopy \cite{ValkunasBook,Cho2005a}. The
second order response, which we have in mind here, corresponds to
the interaction of the excitonic system with first two pulses in a
standard FWM spectroscopy (see Fig. \ref{fig:FWM-Experiment}), and
it should correspondingly describe the time evolution of the system
in the excited or ground state. The part of the response associated
with evolution in the excited state also corresponds to the material
quantities that govern excitation of molecular systems by arbitrary
quantum light \cite{Mancal2010a}.

For a second order response we need to consider only single exciton
band, i.e. the collective excited states of the aggregate with only
a single aggregate member excited. The Frenkel Hamiltonian including
environmental contributions has the general form $H=H_{S}+H_{B}+H_{S-B},$
where the purely electronic (system) Hamiltonian $H_{S}$ reads
\[
H_{S}\equiv\varepsilon^{g}\,|g\rangle\langle g|+\sum_{n}\left(\varepsilon^{n}+\langle\Phi_{n}^{e}-\Phi^{g}\rangle\right)\,|n\rangle\langle n|
\]
\begin{equation}
+\sum_{mn}j_{mn}\,|m\rangle\langle n|,\label{eq:HS}
\end{equation}
the bath Hamiltonian $H_{B}$ can be identified with 
\begin{equation}
H_{B}\equiv T+\Phi^{g},\label{eq:HB}
\end{equation}
and the system-bath interaction $H_{S-B}$ reads as
\begin{equation}
H_{S-B}\equiv\sum_{n}\Delta\Phi_{n}\,|n\rangle\langle n|.\label{eq:HSB}
\end{equation}
Here, $|g\rangle$ is the electronic ground state of the aggregate
of molecules with energy $\varepsilon^{g}$, $|m\rangle$ are electronic
states with the $m^{th}$ molecule of the aggregate of $N$ molecules
excited with energy $\varepsilon^{n}$, $j_{nm}$ represents resonance
coupling between excited states $|m\rangle$ and $|n\rangle$, $\Phi$
represents potential energy surfaces (PES) of the bath DOF in corresponding
electronic states, and $T$ is the kinetic energy of the bath DOF.
We also defined so-called energy gap operator 
\begin{equation}
\Delta\Phi_{n}\equiv\Phi_{n}^{e}-\Phi^{g}-\langle\Phi_{n}^{e}-\Phi^{g}\rangle,\label{eq:EGop}
\end{equation}
where $\langle\bullet\rangle$ represents the equilibrium quantum
mechanical average over the bath DOF. Symbol $\bullet$ denotes an
arbitrary operator. The interaction of the system with light will
be described by the usual semi-classical light-matter interaction
term in dipole approximation
\begin{equation}
H_{S-E}(t)=-\mathbf{\bm{\mu}}\cdot\mathbf{E}(t).\label{eq:System-Light}
\end{equation}
Here, $\bm{\mathbf{\mu}}$ is the transition dipole moment operator
of the aggregate system and $\mathbf{E}(t)=\mathbf{e}E(t)$ is the
electric field vector of the light with polarization vector $\mathbf{e}$. 

The states $|m\rangle$ with excitations localized on individual member
molecules of the aggregate allow us to define the properties of the
aggregate based on crystal structure and quantum chemical inputs.
It is, however, often more convenient to work in a basis of the eigenstate
$|\tilde{m}\rangle$ of the electronic Hamiltonian $H_{S}$. For later
convenience we will define projection operators $K_{m}=|m\rangle\langle m|$
projecting on a state in the local basis and $K_{\tilde{m}}=|\tilde{m}\rangle\langle\tilde{m}|$
projecting on the eigenstate (excitonic) basis. For the discussion
of the non-linear response functions, it is advantageous to use a
superoperator formalism. We define the dipole moment superoperator
as 

\begin{equation}
\mathbf{{\cal V}}\bullet=\frac{1}{\hbar}\left[\mathbf{\mu},\bullet\right]_{-}=\frac{1}{\hbar}[\boldsymbol{\mathbf{\mu}}\cdot\bm{e},\bullet]_{-},\label{eq:Vsup}
\end{equation}
and the Liouvillian as
\begin{equation}
{\cal L}\bullet=\frac{1}{\hbar}[H,\bullet]_{-}.\label{eq:Lsup}
\end{equation}
We will also use the evolution operator
\begin{equation}
U(t)=\exp\left\{ -\frac{i}{\hbar}Ht\right\} ,\label{eq:EvolOp}
\end{equation}
and the evolution superoperator 
\begin{equation}
{\cal U}(t)\bullet=U(t)\bullet U^{\dagger}(t),\label{eq:EvolSup}
\end{equation}
where convenient.

In the second order system-bath coupling theories, the energy gap
operators, Eq. (\ref{eq:EGop}), are present in the equations via
the EGCFs 
\begin{equation}
C_{mn}(t)\equiv\langle U^{\dagger}(t)\Delta\Phi_{m}U(t)\Delta\Phi_{n}\rangle.\label{eq:correlation-function-def}
\end{equation}
Double integration of $C_{mn}(t)$ over time yields so-called \textit{\emph{lineshape
functions}}
\begin{equation}
g_{mn}(t)\equiv\frac{1}{\hbar^{2}}\int\limits _{0}^{t}d\tau\int\limits _{0}^{\tau}d\tau'\; C_{mn}(\tau'),\label{eq:g-function-definition}
\end{equation}
which determine absorption and emission line shapes. Later in the
text, it will be convenient to define the energy gap operators in
the excitonic basis
\begin{equation}
\Delta\Phi_{\tilde{a}\tilde{b}}=\sum_{k}\Delta\Phi_{k}\langle\tilde{a}|k\rangle\langle k|\tilde{b}\rangle,\label{eq:EGopExc}
\end{equation}
the excitonic EGCFs
\begin{align}
C_{\tilde{a}\tilde{b}}(t) & \equiv\langle U^{\dagger}(t)\Delta\Phi_{\tilde{a}\tilde{a}}U(t)\Delta\Phi_{\tilde{b}\tilde{b}}\rangle\\
 & =\sum_{kl}|\langle\tilde{a}|k\rangle|^{2}|\langle\tilde{b}|l\rangle|^{2}C_{kl}(t),
\end{align}
and the excitonic lineshape functions
\begin{equation}
g_{\tilde{a}\tilde{b}}(t)\equiv\frac{1}{\hbar^{2}}\int\limits _{0}^{t}d\tau\int\limits _{0}^{\tau}d\tau'\; C_{\tilde{a}\tilde{b}}(\tau').
\end{equation}

We are interested in the influence of the bath DOF on the dynamics
of the electronic DOF in the excited state. The bath can consists
of several types of collective modes which can be described by various
types of bath correlation function. Experimental situation is typically
well described by one or several overdamped Brownian oscillator modes
standing for an macroscopic number of harmonic DOF, and several underdamped
oscillator modes standing for some important vibrational coordinates,
e.g. normal modes of the chromophore molecules \cite{Nemeth2008a}.
Both of these limits can be conveniently described by by the general
Brownian oscillator correlation function $C(t)$ (see \cite{MukamelBook})
which is a function of temperature $T$, frequency $\Omega_{\mathrm{Bath}}$of
the oscillator, damping coefficient $\gamma$ and the reorganization
energy $\lambda$. In the case of a strongly overdamped mode i.e.
when $\gamma\gg2\Omega_{{\rm bath}}$ the formula simplifies significantly.
At high temperatures we have
\[
C(t)=\hbar\lambda\Lambda\cot(\Lambda\hbar\beta/2)\exp(-\Lambda t)
\]
 
\begin{equation}
-i\hbar\lambda\Lambda\exp(-\Lambda t),
\end{equation}
with $\Lambda=1/\gamma$ . In the opposite case of a non-damped oscillator
$\gamma\rightarrow0$ the correlation function reads
\[
C(t)=\lambda\Omega_{\mathrm{Bath}}\hbar\Big[\coth(\beta\hbar\Omega_{\mathrm{Bath}}/2)\cos(\Omega_{\mathrm{Bath}}t)
\]
 
\begin{equation}
-i\sin(\Omega_{\mathrm{Bath}}t)\Big].\label{eq:Cosc}
\end{equation}

The Hamiltonian, Eq. (\ref{eq:HS}), has a block form. In case we
would consider excited states with up to $N$ excitations per aggregate
a more general Hamiltonian would have to be written including $N+1$
blocks with one block containing just the ground state $|g\rangle$,
and the $N$ blocks containing the excited states with one, two, $\dots$,
up to $N$ excitations. The block structure is enabled by the fact
that the Hamiltonian operators, Eqs. (\ref{eq:HS}) to (\ref{eq:HSB}),
do not contain any terms that enable de-excitation of excited states.
This is a well justified assumption in studies of ultra-fast dynamics
of the chlorophyll based photosynthetic aggregates \cite{Grimm2006}.
The transitions between different blocks of the Hamiltonian are only
enabled by the interaction with the light, Eq. (\ref{eq:System-Light}).
The second order response, which is of interest in this paper, requires
on the ground- and the single-exciton blocks. We will therefore use
upper indices $g$ (ground state) and $e$ (single exciton block)
to denote different blocks of operators and superoperators whenever
we are interested in operations on and evolution of their individual
blocks. Thus, e.g. the time evolution of the system in excited state
single exciton manifold is described by the RDM block $\rho^{(ee)}(t)={\cal U}^{(eeee)}(t)\rho^{(ee)}(0)=U^{(ee)}(t)\rho^{(ee)}(0)U^{(ee)\dagger}(t)$,
and the action of the dipole operator (superoperator) promotes the
ground state block into a coherence block as $\rho^{(eg)}(t)=\mu^{(eg)}\rho^{(gg)}(t)={\cal V}^{(eggg)}\rho^{(gg)}(t)$.
When working in a particular basis of states (e.g. $\{|m\rangle\}=\{|1\rangle,|2\rangle,\dots\}$)
we can write out the sums over the states explicitly, e.g. as
\[
\rho_{mn}^{(ee)}(t)=\sum_{kl}{\cal U}_{mnkl}^{(eeee)}(t)\rho_{kl}^{(ee)}(0)
\]
\begin{equation}
=\sum_{kl}U_{mk}^{(ee)}(t)\rho_{kl}^{(ee)}(0)U_{ln}^{(ee)\dagger}(t).\label{eq:Example}
\end{equation}
The rules are rooted in the simple well-known fact that matrices can
be multiplied by blocks. When discussing the most common non-linear
experiments, the block structure has to be considered up to the two-exciton
block.

\section{Multi-point Correlation Functions in Non-linear Spectroscopy\label{sec:Multi-point-Correlation-Function}}

Non-linear response functions have in general the form of multi-point
correlation functions. For a two band system (ground state and single
excitons), the third order non-linear spectroscopy is completely described
by four response functions \cite{MukamelBook}, listed in Appendix
\ref{sec:Third-Order-Non-linear-Response}. As an example, we will
consider the response usually denoted as $R_{2}$, Eq. (\ref{eq:R2-1}).
We can notice that the block (upper) indices of the evolution operators
in Eq. (\ref{eq:R2-1}) (read from left to right) follow the double-sided
Feynman diagram in Fig. \ref{fig:Double-sided-Feynman-diagrams}A
(see \cite{MukamelBook} for details). During the so-called population
interval $T$ of the response, the system evolves in the excited state
band $(ee)$. If no resonance coupling is present ($j_{mn}=0$ in
Eq. (\ref{eq:HS})) each response function can be split into a sum
of independent components. For $R_{2}$ this means
\begin{equation}
R_{2}(t,T,\tau)=\sum_{kl}R_{2,kl}(t,T,\tau),\label{eq:R2_split}
\end{equation}
where the components
\[
R_{2,kl}(t,T,\tau)=|\mu_{kg}|^{2}|\mu_{lg}|^{2}
\]
 
\begin{equation}
\times{\rm tr}_{B}\{{\cal U}_{kgkg}^{(egeg)}(t){\cal U}_{klkl}^{(eeee)}(T){\cal U}_{glgl}^{(gege)}(\tau)W_{\mathrm{eq}}^{(gg)}\}\label{eq:R2_kl}
\end{equation}
can be evaluated analytically in terms of the line shape functions,
Eq. (\ref{eq:g-function-definition}). Here, $W_{{\rm eq}}^{(gg)}=w_{{\rm eq}}|g\rangle\langle g|$
is the total equilibrium density operator (the electronic energy gap
is assumed to be much higher than $k_{B}T$) and $w_{{\rm eq}}$ represents
the equilibrium density operator of the bath alone. For $j_{mn}>0$
no analytical result is available, and one needs to resort to some
master equation simulations of the evolution superoperators ${\cal U}$.
However, the master equations can only deliver certain reduced (i.e.
averaged over the state of the bath) version of this superoperator. 

The usual way of deriving master equations for the RDM is to apply
a projection operator ${\cal P}$ which reduces the full density matrix
$W(t)$ of the system to its selected part \cite{MayKuehnBook}, in
our case, to the electronic DOF
\begin{equation}
\rho(t)={\rm tr}_{B}\{W(t)\}.\label{eq:RDM}
\end{equation}
The most popular prescription, the so-called Argyres-Kelly (AK) projector
\cite{Argyres1964a}, reads as
\begin{equation}
{\cal P}_{0}W(t)=\rho(t)w_{\mathrm{eq}}\;.\label{eq:AK_proj}
\end{equation}
 It is easy to see that inserting the identity $1={\cal P}+{\cal Q}$,
where ${\cal Q}=1-{\cal P}$ into Eq. (\ref{eq:RDM}) leads to 
\begin{equation}
\rho(t)={\rm tr}_{B}\{{\cal U}(t){\cal P}W(0)+{\cal U}(t){\cal Q}W(0)\}.\label{eq:rho_sep}
\end{equation}
Provided that ${\cal Q}W(0)=0$ (a condition satisfied in the first
interval of the response function, Eq. (\ref{eq:R2-1})) one can define
the reduced evolution superoperator $\tilde{{\cal U}}(t)={\rm tr}_{B}\{{\cal U}(t)w_{{\rm eq}}\}$
such that $\rho(t)=\tilde{{\cal U}}(t)\rho(0)$. The evolution superoperator
$\tilde{{\cal U}}(t)$ can be calculated from a master equation that
can be derived by projection operator formalism \cite{MayKuehnBook}. 

Let us apply the same method to higher order response functions. Let
use assume we have derived a master equation by applying the corresponding
projector operator ${\cal P}$ and calculated elements of the reduced
evolution superoperator $\tilde{{\cal U}}(t)$. We can then assemble
an approximate response function
\[
\tilde{R}_{2,kl}(t,T,\tau)=\tilde{{\cal U}}_{kgkg}^{(egeg)}(t)
\]
 
\begin{equation}
\times\tilde{{\cal U}}_{klkl}^{(eeee)}(T)\tilde{{\cal U}}_{glgl}^{(gege)}(\tau)\rho^{(gg)}(0).\label{eq:R2_tilde}
\end{equation}
Here, we set all transition dipole moment elements to one for the
sake of brevity. It can be shown that $\tilde{R}_{2,kl}$ can also
be written as
\[
\tilde{R}_{2,kl}(t,T,\tau)={\rm tr}_{B}\{{\cal U}_{kgkg}^{(egeg)}(t)
\]
 
\begin{equation}
\times{\cal P}{\cal U}_{klkl}^{(eeee)}(T){\cal P}{\cal U}_{glgl}^{(gege)}(\tau)\rho^{(gg)}(0)\}.\label{eq:R2_tilde_2}
\end{equation}
However, the exact expression for $R_{2,kl}$ reads as 
\[
R_{2,kl}(t,T,\tau)={\rm tr}_{B}\{{\cal U}_{kgkg}^{(egeg)}(t)({\cal P}+{\cal Q}){\cal U}_{klkl}^{(eeee)}(T)
\]
\begin{equation}
\times({\cal P}+{\cal Q}){\cal U}_{glgl}^{(gege)}(\tau)\rho^{(gg)}(0)\}.\label{eq:R2_still_exa}
\end{equation}
Eqs. (\ref{eq:R2_tilde_2}) and (\ref{eq:R2_still_exa}) differ by
${\cal Q}$-containing terms that cannot be in general eliminated,
and consequently one cannot expect master equations based on a single
projector operator ${\cal P}$ of any type to reproduce the third
order response functions. This applies also to the validity of non-perturbative
schemes of calculations of non-linear response, such as those derived
in Refs. \cite{Gelin2005a} and \cite{Mancal2006a}. 

A general solution of this problem was proposed in Ref. \cite{Mancal2010c}.
It was argued that one cannot write down a single exact master equation
for all three intervals of the response function. Rather, one has
to write a different master equation for each interval. This is formally
possible by introducing three different projectors ${\cal P}_{0}$
(i.e. standard AK projector) for the first, $P_{\tau}$ for the second
and $P_{\tau+T}$ for the third interval of the response \cite{Mancal2010c}.
The projectors $P_{\tau}$ and $P_{\tau+T}$ are constructed so as
to cancel the ${\cal Q}$-containing term exactly for $j_{mn}=0$.
In this limit, all response functions, Eqs. (\ref{eq:R1-1}) to (\ref{eq:R4-1}),
can be exactly reproduced by the corresponding master equations. In
the following sections, we will treat the case of $j_{mn}\neq0$,
for the case of the second order response, i.e. we will derive equations
of motion for the reduced density matrix using the projector $P_{\tau}$.
The application of this approach to the full response function will
be treated elsewhere. An alternative to the projector approach is
to attempt to derive specific equations of motion for each response
function as a whole in a specific perturbation scheme, e.g. second
order convolutionless QME (CL-QME) as in Ref. \cite{Richter2010a}.
While both approaches should lead to similar results, the projection
operator technique is not limited to any specific way of expanding
the equations of motion in terms of system-bath coupling, and we believe
it is therefore somewhat more flexible.

\section{Second Order Response to Light\label{sec:Second-Order-Response}}

The second order optical response can yield an optical signal on the
sum or difference frequencies of the exciting field. Here, we consider
only the degenerate experiment when the excitation field have the
same frequency, and we are interested in particular in the zero frequency
part of the response which does not generate an optical signal. Unlike
in spectroscopy where we study a quantum mechanical expectation value
(polarization or field), here we would like to study the state of
the system achieved by the excitation. Correspondingly, the second
order response will not be expressed in terms of response function,
but rather in terms of some response operators. By the {}``system''
we mean electronic DOF, and thus the response operators correspond
to the reduced density matrix of the system in the same way the response
function corresponds to an expectation value of e.g. polarization.

\subsection{Excited State of a Molecular System}

Assuming that we excited the system by some external field $E(t)$,
the second order reduced density operator of the system reads as
\[
\hat{\rho}^{(2)}(t)=-\int\limits _{0}^{\infty}{\rm d}t_{1}\int\limits _{0}^{\infty}{\rm d}t_{2}{\rm tr}_{B}\{{\cal U}(t_{2}){\cal V}
\]
\begin{equation}
\times{\cal U}(t_{1}){\cal V}W_{{\rm eq}}|g\rangle\langle g|\}E(t-t_{2})E(t-t_{2}-t_{1}).
\end{equation}
The excited state part of the second order density operator $\hat{\rho}^{(2)}$
then yields
\[
\rho^{(2)}(t)=
\]
\[
\int\limits _{0}^{\infty}{\rm d}t_{1}\int\limits _{0}^{\infty}{\rm d}t_{2}\Big[R_{I}(t_{2},t_{1})A^{*}(t-t_{2})A(t-t_{2}-t_{1})e^{i\omega t_{1}}
\]
\begin{equation}
+R_{II}(t_{2},t_{1})A(t-t_{2})A^{*}(t-t_{2}-t_{1})e^{-i\omega t_{1}}\Big],\label{eq:rho_ee_pulse}
\end{equation}
where
\[
R_{I}(t_{2},t_{1})=\mathrm{Tr}{}_{{\rm B}}\Big\{{\cal U}^{(eeee)}(t_{2})
\]
\begin{equation}
\times{\cal V}^{(eeeg)}{\cal U}^{(egeg)}(t_{1}){\cal V}^{(eggg)}w_{\mathrm{eq}}|g\rangle\langle g|\Big\},\label{eq:R_I}
\end{equation}
and
\[
R_{II}(t_{2},t_{1})=\mathrm{Tr}{}_{{\rm B}}\Big\{{\cal U}^{(eeee)}(t_{2})
\]
\begin{equation}
\times{\cal V}^{(eege)}{\cal U}^{(gege)}(t_{1}){\cal V}^{(gegg)}w_{\mathrm{eq}}|g\rangle\langle g|\Big\},\label{eq:R_II}
\end{equation}
are the second order response operators. In Eq. (\ref{eq:rho_ee_pulse})
we introduced the electric field envelopes $A(t)$ and the carrier
frequency $\omega$ so that $E(t)=A(t)e^{-i\omega t}+c.c.$. The interaction
of the molecular system with arbitrary light can be expressed using
the operators $R_{I}$ and $R_{II}$ as discussed in Ref. \cite{Mancal2010a}.
For a general state of the light $|\psi_{{\rm rad}}\rangle$, the
terms $A^{*}(t-t_{2})A(t-t_{2}-t_{1})e^{i\omega t_{1}}$ have to be
replaced by the light correlation function $I(t-t_{2},t-t_{2}-t_{1})=\langle\psi_{{\rm rad}}|\hat{E}^{\dagger}(t-t_{2})\hat{E}(t-t_{2}-t_{1})|\psi_{{\rm rad}}\rangle$,
where $\hat{E}(t)$ is the operator of the electric field in Heisenberg
representation. This allows us to calculate the state of a system
excited by light with arbitrary properties. 

In this paper, we will concentrate on the situation when the molecular
system is excited by two ultra-short laser pulses traveling in two
different directions $\bm{k}_{1}$ and $\bm{k}_{2}$ of which the
second one arrives with a delay $\tau$. Thus we have $A_{1}(t)=A_{0}e^{i\bm{k}_{1}\cdot\bm{r}}\delta(t+\tau)$,
$A_{2}(t)=A_{0}e^{i\bm{k}_{2}\cdot\bm{r}}\delta(t)$. The time zero
is set to the center of the second pulse. This corresponds to a typical
situation in the coherent non-linear spectroscopy, and the time $t$
in which the excited state evolves corresponds to the population time
$T$ of the non-linear spectroscopy. Inserting the delta function
envelopes into Eq. (\ref{eq:rho_ee_pulse}) and assuming that the
pulse with wave-vector $\bm{k}_{1}$ precedes the pulse with wave-vector
$\bm{k}_{2}$, we arrive at
\begin{equation}
\hat{\rho}_{I}^{(2)}(t;\tau)=R_{I}(t,\tau)|A_{0}|^{2}.\label{eq:rhoee_I}
\end{equation}
 This operator corresponds to the excited state time evolution in
the left hand side (l.h.s.) diagram of Fig. \ref{fig:Double-sided-Feynman-diagrams}B.
The base of this diagram corresponds to the so-called rephasing pathways.
When the pulse sequence is reverted (pulse $\bm{k}_{1}$ arrives second
with delay $\tau$) we obtain so-called non-rephasing pathways and
the time evolution corresponds to the operator
\begin{equation}
\hat{\rho}_{II}^{(2)}(t;\tau)=R_{II}(t,\tau)|A_{0}|^{2}.\label{eq:rhoee_II}
\end{equation}

It is easy to verify that $\hat{\rho}_{I}^{(2)}(t;\tau)$ and $\hat{\rho}_{I}^{(2)}(t;\tau)$
do not have to be Hermitian, and they alone cannot be said to represent
a state of a molecular system. This is the result of them being only
a portion of the perturbation expansion of the non-linear response
operator. Only the sum of Eqs. (\ref{eq:R_I}) and (\ref{eq:R_II})
yields a Hermitian operator which can describe excited state of a
molecular system. This situation has an analogy in the case of 2D
coherent spectroscopy where the rephasing and non-rephasing signals
alone cannot be interpreted as representing absorption and emission
events, while the sum spectrum can be assigned this interpretation
\cite{Jonas2003a}. When the system is excited by a single finite
length pulse, the two contributions to the excited state are equally
weighted, guaranteeing thus the proper properties of the corresponding
density matrix. 

In the following section we will discuss the equations of motion for
$\hat{\rho}_{I}^{(2)}(t;\tau)$ and $\hat{\rho}_{II}^{(2)}(t;\tau)$
to asses the influence of the delay $\tau$ on the dynamics of their
diagonal (populations) and off-diagonal (coherence) elements. We will
thus attempt to answer the question whether in the non-linear spectroscopic
methods, such as 2D coherent spectroscopy, we observe the expected
excited state dynamics, i.e. the one unaffected by the delay $\tau$.

\subsection{Simulation of the Response by Reduced Density Matrix Equations}

In order to calculate the second order non-linear response, we need
to evaluate the component expressions of Eq. (\ref{eq:R_II}). We
start with equations of motion for the perturbation of the total density
matrix $W(t)$. We define the first order operator as
\begin{equation}
W_{II}^{(1)}(t)={\cal U}^{(gege)}(t){\cal V}^{(gegg)}w_{\mathrm{eq}}|g\rangle\langle g|.\label{eq:W_1_II}
\end{equation}
This operator corresponds to the evolution after the first interaction
of the system with light in response function, Eq. (\ref{eq:R_II}).
The response function has to be read from left to right. The operator
$W_{II}^{(1)}(t)$ satisfies the following equation 
\begin{equation}
\frac{\partial}{\partial t}W_{II}^{(1)}(t)=-i{\cal L}^{(gege)}W_{II}^{(1)}(t),\label{eq:W_1_II_eq}
\end{equation}
with the initial condition $W_{II}^{(1)}(t=0)={\cal V}^{(gegg)}w_{{\rm eq}}|g\rangle\langle g|=w_{{\rm eq}}|g\rangle\langle g|\mu^{(ge)}$.
RDM equation of motion can be found applying standard AK projection
operator, Eq. (\ref{eq:AK_proj}), for which $(1-{\cal P}_{0})W_{II}^{(1)}(t=0)$
is equal to zero identically. The procedure of the derivation of the
RDM equation of motion leads to
\begin{equation}
\frac{\partial}{\partial t}\rho_{II}^{(1)}(t)=-i\bar{{\cal L}}_{0}^{(gege)}\rho_{II}^{(1)}(t)-{\cal R}_{0}^{(gege)}(t)\rho_{II}^{(1)}(t),\label{eq:RDM_1_II}
\end{equation}
where $\bar{{\cal L}}_{0}^{(gege)}$ is the coherence block of the
electronic Liouville superoperator, and ${\cal R}_{0}^{(gege)}(t)$
is some second order dephasing tensor. The particular form of ${\cal R}_{0}^{(gege)}$
depends on the approximations and the theory applied. For a pure dephasing
model and a harmonic bath, the dephasing tensor which follows from
a second order CL-QME (see e.g. \cite{Olsina2010a}) can be shown
to yield an exact dynamics for $\rho_{II}^{(1)}(t)$ \cite{Doll2008a}.
In secular approximation, Eq. (\ref{eq:RDM_1_II}) yields
\begin{equation}
\frac{\partial}{\partial t}\rho_{II,\tilde{k}g}^{(1)}(t)=-i\omega_{kg}\rho_{II,\tilde{k}g}^{(1)}(t)-\int_{0}^{t}{\rm d}\tau C_{\tilde{k}\tilde{k}}(\tau)\rho_{II,\tilde{k}g}^{(1)}(t),\label{eq:I_interv}
\end{equation}
which is a set of independent equations for optical coherences. Note
that we use index $\tilde{k}$ to denote the excitonic representation.
In the limit of $j_{mn}\rightarrow0$ (pure dephasing) the states
$|\tilde{k}\rangle$ and $|k\rangle$ coincide.

The evolution in the second propagation interval can be expressed
through the density operator
\begin{equation}
W_{II}^{(2)}(t;\tau)={\cal U}^{(eeee)}(t){\cal V}^{(eeeg)}W_{II}^{(1)}(\tau),\label{eq:W_2_II}
\end{equation}
which satisfies
\begin{equation}
\frac{\partial}{\partial t}W_{II}^{(2)}(t;\tau)=-i{\cal L}^{(eeee)}W_{II}^{(2)}(t;\tau),\label{eq:W_2_II_eq}
\end{equation}
with the initial condition $W_{II}^{(2)}(t=0;\tau)={\cal V}^{(eege)}W_{II}^{(1)}(\tau)=\mu^{(eg)}W_{II}^{(1)}(\tau)$.
The application of the AK projector would lead to a loss of information
in the construction of the response function, because the term ${\cal Q}_{0}\mu^{(eg)}W^{(1)}(\tau)=(1-{\cal P}_{0})W^{(1)}(\tau)\neq0$.
Consequently, even if we successfully derive an exact master equation
for $\rho_{II}^{(2)}(t)=\mathrm{Tr_{B}}\{W_{II}^{(2)}(t)\}$, the
response function constructed from this equation would be missing
the ${\cal Q}$-containing terms. The parametric projection operator
recently suggested in Ref. \cite{Mancal2010c} has the property of
eliminating the initial term approximately, $(1-{\cal P}_{\tau})\mu_{eg}W_{II}^{(1)}(\tau)\approx0$
. In the case of pure dephasing, it turns to zero exactly. We can
therefore write an equation of motion for $\rho_{II}^{(2)}$ which
is analogical to Eq. (\ref{eq:RDM_1_II}) 
\[
\frac{\partial}{\partial t}\rho_{II}^{(2)}(t;\tau)=-i\bar{{\cal L}}_{\tau}^{(eeee)}\rho_{II}^{(2)}(t;\tau)
\]
\begin{equation}
-{\cal R}_{\tau}^{(eeee)}(t)\rho_{II}^{(2)}(t;\tau).\label{eq:RDM_2_II}
\end{equation}
Again, the $\bar{{\cal L}}_{\tau}^{(eeee)}$ is the electronic Liouville
superoperator and ${\cal R}_{\tau}^{(eeee)}(t)$ is the superoperator
describing the electronic energy relaxation and the electronic coherence
dephasing in the single-exciton band. In the following section, we
will derive the relaxation tensor ${\cal R}_{\tau}^{(eeee)}(t)$ corresponding
to the projection operator ${\cal P}_{\tau}$.

\section{master equations with parametric projector\label{sec:Reduced-density-matrix} }

In this section, we will apply the well-known Nakajima-Zwanzig identity
in the second order in system bath interaction Hamiltonian together
with the parametric projection operator proposed in Ref. \cite{Mancal2010c}.
We are interested in the second interval of the non-linear response,
and in the evolution in the single exciton band in particular .

\subsection{The Choice of Projector\label{sub:Nakajima-Zwanzig-Identity-and}}

\begin{figure}[b]
\centering{}\includegraphics[width=1\columnwidth]{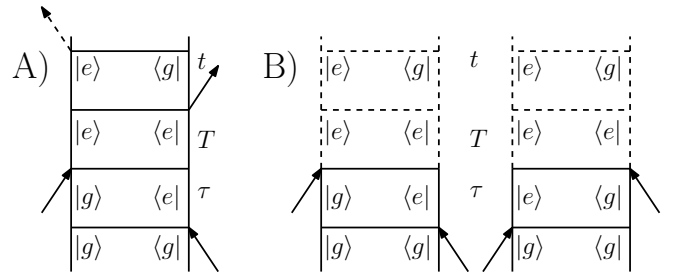} \caption{\label{fig:Double-sided-Feynman-diagrams}Double-sided Feynman diagrams
of the third and the second order response. Part A: The Feynman diagram
of the Liouville pathway $R_{2}$. Part B: In full lines, the Feynman
diagrams of the response operators $R_{II}$ (left) and $R_{I}$ (right).
The dashed part completes the diagram into a corresponding third order
response.}

\end{figure}

The projector $\mathcal{P}_{\tau}$ is chosen in such way that it
contains time evolution of the bath in the first interval of the response,
where the relevant system dynamics is the one of an optical coherence.
The projector corresponding to the pathway $R_{II}$ and the length
of the first interval $\tau$ reads according to Ref. \cite{Mancal2010c}
\begin{equation}
\mathcal{P}_{\tau,II}\bullet=\mathrm{tr}_{B}\left\{ \bullet\right\} \, U^{g}(\tau)w_{\mathrm{eq}}\,\sum_{\tilde{n}}U_{\tilde{n}}^{e\dagger}(\tau)\, K_{\tilde{n}}\, e^{g_{\tilde{n}\tilde{n}}^{*}(\tau)}\;.\label{eq:projector-definition2}
\end{equation}
This choice gives an exact description of the bath for $j_{mn}=0$.
For a non-zero coupling it corresponds to the secular approximation
in the first interval equation of motion, Eq. (\ref{eq:I_interv}).
Further on in the text, all derivations will be done for the pathway
$R_{II}$, and we omit the index $II$. The treatment of the pathway
$R_{I}$ is analogical. We define evolution operators describing the
dynamics of the environmental DOF while the system is in its electronic
ground state

\begin{equation}
U^{g}(\tau)=\exp\left(-\frac{i}{\hbar}(T+\Phi^{g})\tau\right),\label{eq:Ug-def}
\end{equation}
and in the excited eigenstate $|\tilde{n}\rangle$
\begin{equation}
U_{\tilde{n}}^{e}(\tau)=\exp\left(-\frac{i}{\hbar}(T+\Phi_{\tilde{n}\tilde{n}}^{e}-\langle\Phi_{\tilde{n}\tilde{n}}^{e}-\Phi^{g}\rangle)\tau\right).\label{eq:Ue-def}
\end{equation}
By the choice of the projector, Eq. (\ref{eq:projector-definition2}),
we prescribe an ansatz 
\begin{equation}
W^{(2)}(0)=W^{(1)(I)}(\tau)\equiv\rho^{(1)(I)}(\tau)w_{\tau}^{(I)},
\end{equation}
where
\begin{equation}
w_{\tau}^{(I)}\equiv\sum_{\tilde{n}}K_{\tilde{n}}U^{g}(\tau)w_{\mathrm{eq}}U_{\tilde{n}}^{e\dagger}(\tau)\; e^{g_{\tilde{n}\tilde{n}}^{*}(\tau)}\;.\label{eq:wItau_def}
\end{equation}
For zero resonance coupling $j_{mn}$, this is an exact prescription
for the bath. For non-zero resonance coupling, it is an approximation
comparable to the secular approximation. Projector, Eq. (\ref{eq:projector-definition2}),
can be written in short as
\begin{align}
\mathcal{P}_{\tau}\bullet & =\mathrm{tr}_{B}\left\{ \bullet\right\} w_{\tau}^{(I)}.\label{eq:projector-druhy-interval}
\end{align}
It is important to note that $w_{\tau}^{(I)}$ is not a purely bath
operator, and it does not generally commute with $\rho$. Also the
interaction picture with respect to the system Hamiltonian $H_{S}$
denoted by $(I)$ applies to it. It stands on the right hand side
(r.h.s.) of $\rho$ when evaluating $R_{II}$, while in $R_{I}$ it
would stand on the l.h.s. of $\rho$. This follows from the diagrams
in Fig. \ref{fig:Double-sided-Feynman-diagrams}B. %
The full form of
$W^{(1)}(\tau)$ is 
\begin{align}
W^{(1)}(\tau)= & \; U^{g}(\tau)w_{\mathrm{eq}}\times\,\nonumber \\
 & \sum_{\tilde{n}}U_{\tilde{n}}^{e\dagger}(\tau)\, K_{\tilde{n}}e^{i(\varepsilon_{\tilde{n}}+\langle\Phi_{\tilde{n}}^{e}-\Phi^{g}\rangle)\tau/\hbar}\rho^{(1)}(0).
\end{align}
From now on, the upper index $(2)$ will be omitted in text for the
sake of brevity. 

We can verify that the projector property $\mathcal{P}_{\tau}^{2}=\mathcal{P}_{\tau}$
is fulfilled 
\begin{align}
\mathcal{P}_{\tau} & \mathcal{P}_{\tau}\left(|\tilde{m}\rangle\langle\tilde{n}|\, w\right)\nonumber \\
 & =|\tilde{m}\rangle\langle\tilde{n}|\,\mathrm{tr}_{B}\left\{ U^{g}(\tau)w_{\mathrm{eq}}\, U_{\tilde{n}}^{e\dagger}(\tau)\right\} \,\mathrm{tr}_{B}\left\{ w\right\} \, e^{2g_{\tilde{n}\tilde{n}}^{*}(\tau)}\nonumber \\
 & =|\tilde{m}\rangle\langle\tilde{n}|\, e^{-g_{\tilde{n}\tilde{n}}^{*}(\tau)}\,\mathrm{tr}_{B}\left\{ w\right\} \, e^{2g_{\tilde{n}\tilde{n}}^{*}(\tau)}=\mathcal{P}_{\tau}\left(|\tilde{m}\rangle\langle\tilde{n}|\, w\right)\;.
\end{align}
Here, we used expression for the line shape function $g(t)$ in the
second cumulant approximation 
\begin{equation}
e^{-g(t)}=\mathrm{tr}_{B}\left\{ U^{g\dagger}(t)U^{e}(t)w_{\mathrm{eq}}\right\} \;.
\end{equation}
The action of the projector $\mathcal{P}_{\tau}(\equiv\mathcal{P}_{\tau,II})$
on the electronic state is asymmetric, because the projector was derived
for the Liouville pathway II (see Fig \ref{fig:Double-sided-Feynman-diagrams}B).

\subsection{Parametric Master Equation}

Now, we apply the projector ${\cal P}_{\tau}$, Eq. (\ref{eq:projector-druhy-interval}),
to the Nakajima-Zwanzig identity in the interaction picture

\begin{align}
 & \frac{\partial}{\partial t}\mathcal{P}_{\tau}W^{(I)}(t)=\mathrm{NZ}_{1}+\mathrm{NZ}_{2}+\mathrm{NZ}_{3}=\nonumber \\
 & \;\;\;\;\;-\mathcal{P}_{\tau}\mathcal{L}^{(I)}(t)\,\exp\left(-i\int\limits _{t_{0}}^{t}d\tau^{\prime}\;\mathcal{Q}_{\tau}\mathcal{L}^{(I)}(\tau^{\prime})\mathcal{Q}_{\tau}\right)\mathcal{Q}_{\tau}W(t_{0})\nonumber \\
 & \;\;\;\;\;-\int\limits _{0}^{t-t_{0}}d\tau'\;\Bigg[\mathcal{P}_{\tau}\mathcal{L}^{(I)}(t)\,\exp\left(-i\int_{t_{0}}^{\tau'}d\tau''\;\mathcal{Q}_{\tau}\mathcal{L}^{(I)}(\tau'')\mathcal{Q}_{\tau}\right)\nonumber \\
 & \;\;\;\;\;\times\mathcal{Q}_{\tau}\mathcal{L}^{(I)}(t-\tau')\mathcal{P}_{\tau}W^{(I)}(t-\tau')\Bigg]\nonumber \\
 & \;\;\;\;\; -i\mathcal{P}_{\tau}\mathcal{L}^{(I)}(t)\mathcal{P}_{\tau}W^{(I)}(t).\label{eq:nakajima-zwanzig}
\end{align}
We will use Eq. (\ref{eq:nakajima-zwanzig}) up to the second order
in $\mathcal{L}^{(I)}$, and we set $t_{0}=0$. The term ${\rm NZ}_{1}$
corresponds to so-called initial term, which we made equal to zero
by the choice of the projector. The last term ${\rm NZ_{3}}$ corresponds
to an effective Liouvillian, and it is usually a purely electronic
operator. Now, with the parametric projector it contains additional
terms originating from the system bath interaction. Its purely electronic
part will stand for the effective Liouvillian $\bar{{\cal L}}_{\tau}^{(eeee)}$
in Eq. (\ref{eq:RDM_2_II}), while its additional $\tau$-depending
contribution we will add to the relaxation tensor ${\cal R}_{\tau}^{(eeee)}$.
Finally, the term ${\rm NZ_{2}}$ is a starting point for the derivation
of the second order relaxation term, which has a form of a convolution
between the RDM and some memory kernel. In the second order expansion
of Eq. (\ref{eq:nakajima-zwanzig}) and with an approximation $W^{(I)}(t-\tau^{\prime})\approx W^{(I)}(t)$,
the resulting equation for ${\cal P}_{\tau}W^{(I)}(t)$ coincides
with the second order approximation of the equivalent time-convolutionless
identity \cite{FainBook}. Thus the ${\rm NZ}_{2}$ will lead to a
contribution to the tensor ${\cal R}_{\tau}^{(eeee)}$ in Eq. (\ref{eq:RDM_2_II}).
However, the parametric projection operator technique allows us to
keep the convolution form of the equation if it is desired. While
the convolution form may have some advantages over the CL-QME \cite{Olsina2010a},
only the CL-QME leads in the limit of $j_{mn}=0$ to a result coinciding
with the one obtained by the second cumulant treatment of the non-linear
response functions (see e.g. Ref. \cite{Mancal2010c} and the Appendix
\ref{sec:Appendix})

Let us point out the most important aspect of the application of the
projector ${\cal P}_{\tau}$ with the identity, Eq. (\ref{eq:nakajima-zwanzig}).
It is important to note that the projector ${\cal P}_{\tau}$ itself
contains the system-bath interaction to all orders in the form of
the exponential of the line-shape function (see Eq. (\ref{eq:projector-definition2})).
The success of the second order master equations (of the form $\dot{\rho}=-\alpha_{2}\rho$,
where the dot denotes the time derivative, and $\alpha_{2}$ is some
second order operator) lies in the fact that their solutions includes
all orders of the perturbation ($\rho=e^{-\alpha_{2}t}\rho_{0}$).
The solution corresponds to a partial summation of the perturbative
series to infinity. For some types of bath, such as the bath consisting
of harmonic oscillators, this may even lead to exact master equations
\cite{Doll2008a}. When higher order terms are added to the right
hand side of the equation motion by a procedure that does not respect
the form of higher order terms dictated by the Nakajima-Zwanzig identity,
the resulting equation of motion may lead to unphysical results. Therefore,
one has to take care in application of the projector ${\cal P}_{\tau}$,
not to allow higher than second order contributions to appear on the
right hand side of Eq. (\ref{eq:nakajima-zwanzig}). Since the difference
between projectors ${\cal P}_{0}$ and $\mathcal{P}_{\tau}$ is only
in dynamics of the system-bath coupling during time $\tau$, their
difference is at least of the first order in $\Delta\Phi$. Since
the $\mathrm{NZ}_{2}$ with AK projector is already of the second
order in $\Delta\Phi$ in all its terms, the difference caused by
using projector $\mathcal{P}_{\tau}$ will be of higher order in $\Delta\Phi$.
In the second order master equation, the term $\mathrm{NZ}_{2}$ with
projector $\mathcal{P}_{\tau}$ has to be equivalent to the form obtained
from with AK projector (see e.g. \cite{MayKuehnBook}). Applying the
following approximation $\rho^{(I)}(t-\tau^{\prime})\approx\rho^{(I)}(t)$
in the term ${\rm NZ}_{2}$ we obtain it in the form the second order
relaxation term of CL-QME \cite{FainBook}.

The details of the evaluation of the term ${\rm NZ}_{3}$ are presented
in Appendix \ref{sec:Derivation-of-Relaxation}. As expected it yields
a $\tau$-dependent term. Putting all results together and tracing
over bath DOF we obtain

\begin{align}
\frac{\partial}{\partial t}\rho^{(I)}(t;\tau) & =\;\sum_{mn}\int\limits _{0}^{t}ds\;\Big[\nonumber \\
\, & -\ddot{g}_{mn}(s)K_{m}(t)K_{n}(t-s)\rho^{(I)}(t)\nonumber \\
\, & +\ddot{g}_{mn}^{*}(s)K_{m}(t)\rho^{(I)}(t)K_{n}(t-s)\nonumber \\
\, & +\ddot{g}_{mn}(s)K_{n}(t-s)\rho^{(I)}(t)K_{m}(t)\nonumber \\
\, & -\ddot{g}_{nm}^{*}(s)\rho^{(I)}(t)K_{n}(t-s)K_{m}(t)\Big]\nonumber \\
+ & \,\sum_{m\tilde{k}}\,\left(K_{m}(t)\rho^{(I)}(t)K_{\tilde{k}}-\rho^{(I)}(t)K_{\tilde{k}}K_{m}(t)\right)\nonumber \\
 & \;\;\;\times\left(\dot{g}_{m\tilde{k}}^{*}(t+\tau)-\dot{g}_{m\tilde{k}}^{*}(t)\right)\;.\label{eq:Final-Int}
\end{align}
The first five lines of Eq. (\ref{eq:Final-Int}) corresponds to the
standard CL-QME, while the last two lines represent the $\tau$-dependent
contribution which the standard CL-QME does not predict.

Let us investigate the $\tau$-dependent term only. First, we turn
to Schr\"{o}dinger picture by substituting $\rho^{(I)}(t)=U_{S}^{\dagger}(t)\rho(t)U_{S}(t)$.
We denote the new $\tau$-dependent term of the CL-QME by ${\cal D}(t;\tau)$
\begin{align}
{\cal D}(t;\tau)\rho(t) & =\nonumber \\
 & \;\sum_{m\tilde{k}}\,\left(K_{m}(t)\rho(t)K_{\tilde{k}}-\rho(t)K_{\tilde{k}}K_{m}(t)\right)\nonumber \\
 & \;\;\;\times\left(\dot{g}_{m\tilde{k}}^{*}(t+\tau)-\dot{g}_{m\tilde{k}}^{*}(t)\right)\;.\label{eq:New-terms-final}
\end{align}
It can be easily verified that the new term preserves the trace of
$\rho(t)$, because $\mathrm{Tr}\left(K_{m}(t)\rho(t)K_{\tilde{k}}-\rho(t)K_{\tilde{k}}K_{m}(t)\right)=0$.
From the inspection of the term $\dot{g}_{m\tilde{k}}^{*}(t+\tau)-\dot{g}_{m\tilde{k}}^{*}(t)$
we can conclude that the effect of the parameter $\tau$ is transient.
If EGCF tends to zero on a time scale given by some bath correlation
time, this term also tends to zero. The dynamics at long times is
therefore not affected by the delay between the two excitation interactions.
In Appendix \ref{sec:Homodimer} we derived the difference term, Eq.
(\ref{eq:New-terms-final}), of a special case of molecular homodimer
with bath fluctuations uncorrelated between the sides. We found the
$\tau$-dependent term to be identically equal to zero in this case.
In the next section we study the effect of the $\tau$-dependent term
numerically for a heterodimer.

\section{Numerical Results and Discussion\label{sec:Numerical-Results-and}}

In this section, we study the dynamics of the elements of the response
functions, Eqs. (\ref{eq:R_I}) and (\ref{eq:R_II}). The RDM $\rho(t;\tau)$
for which we derived Eq. (\ref{eq:Final-Int}) corresponds the response
function $R_{II}$ (see Eq. (\ref{eq:rhoee_II}) for the case of $A_{0}=1$).
We compare the dynamics in two cases. In the first case, the evolution
operator ${\cal U}^{(eeee)}(T)$ appearing in Eqs. (\ref{eq:R_I})
and (\ref{eq:R_II}) is calculated by standard time dependent CL-QME
derived using AK projector, Eq. (\ref{eq:AK_proj}). In the second
case, it is calculated using Eq. (\ref{eq:Final-Int}). We use the
relation $\rho_{I,ij}(t)=\rho_{II,ji}^{*}(t)$ between the RDMs from
different Liouville pathways.

\begin{figure}[b]
\centering{}\includegraphics[clip,width=0.6\columnwidth]{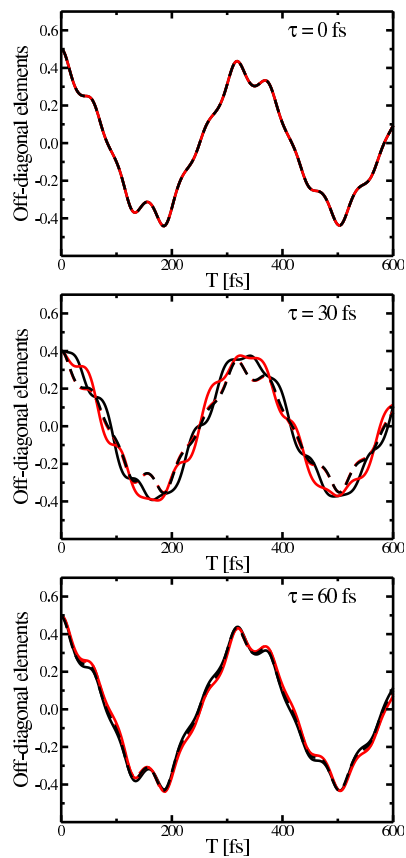} \caption{The time evolution of the real parts of $R_{I}$ and $R_{II}$ for
a heterodimer with energy gap $\Delta=100\;\mathrm{cm}^{\mbox{-}1}$,
interacting with a bath represented by a general Brownian oscillator
with parameters $\lambda=50\;\mathrm{cm^{\mbox{-}1}}$, $\gamma=1\;\mathrm{ps^{\mbox{-}1}}$and
$\Omega_{\mathrm{Bath}}=100\;\mathrm{ps^{\mbox{-}1}}$. Red lines,
the matrix element $12$ of $R_{I}$ according to standard CL-QME
(dashed) and the corresponding parametric CL-QME (full). Black lines,
the matrix element $12$ of $R_{II}$ according to standard CL-QME
(dashed) and the corresponding parametric CL-QME (full).\label{fig:1}}

\end{figure}

\begin{figure}[h]
\centering{}\includegraphics[clip,width=0.6\columnwidth]{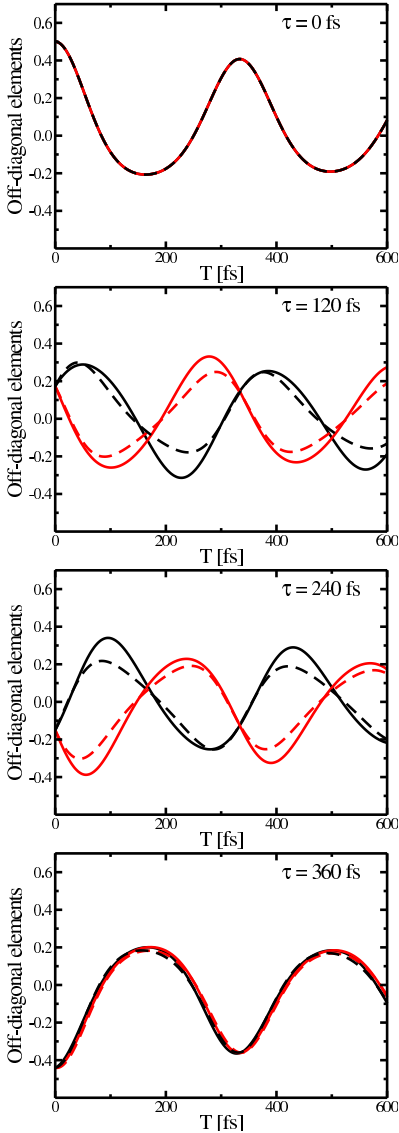} \caption{The time evolution of the real parts of $R_{I}$ and $R_{II}$ for
a heterodimer with energy gap $\Delta=100\;\mathrm{cm}^{\mbox{-}1}$,
interacting with a bath represented by a general Brownian oscillator
with parameters$\lambda=5\;\mathrm{cm^{\mbox{-}1}}$, $\gamma=1\;\mathrm{ps^{\mbox{-}1}}$
and $\Omega_{\mathrm{Bath}}=19\;\mathrm{ps^{\mbox{-}1}}$. Red lines,
the matrix element $12$ of $R_{I}$ according to standard CL-QME
(dashed) and the corresponding parametric CL-QME (full). Black lines,
the matrix element $12$ of $R_{II}$ according to standard CL-QME
(dashed) and the corresponding parametric CL-QME (full).\label{fig:2}}

\end{figure}

\begin{figure}[h]
\centering{}\includegraphics[clip,width=1\columnwidth]{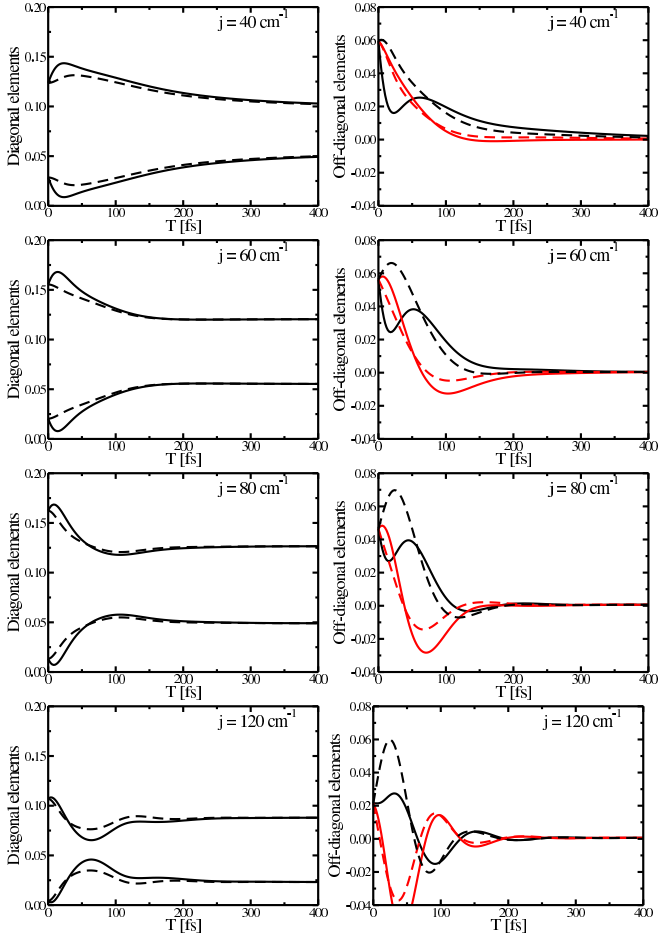} \caption{The time evolution of the real parts of $R_{I}$ and $R_{II}$ and
their dependency on the resonance coupling $j$ for a heterodimer
with energy gap $\Delta=100\;\mathrm{cm}^{\mbox{-}1}$ and an anti-parallel
arrangement of the transition dipole moments, interacting with a bath
represented by an overdamped Brownian oscillator with reorganization
energy $\lambda=120\;\mathrm{cm^{\mbox{-}1}}$ and correlation time
$\tau_{c}=\Lambda^{-1}=50\;\mathrm{fs}$. The delay between the two
pulses is fixed to $\tau=60\;\mathrm{fs}$. Black lines, the matrix
elements $11$, $22$ (left column) and $12$ (right column) of $R_{II}$
according to standard CL-QME (dashed) and the corresponding parametric
CL-QME (full). Red lines, the matrix element $12$ (right column)
of $R_{I}$ according to standard CL-QME (dashed) and the corresponding
parametric CL-QME (full).\label{fig:3}}

\end{figure}

\begin{figure}[h]
\centering{}\includegraphics[clip,width=1\columnwidth]{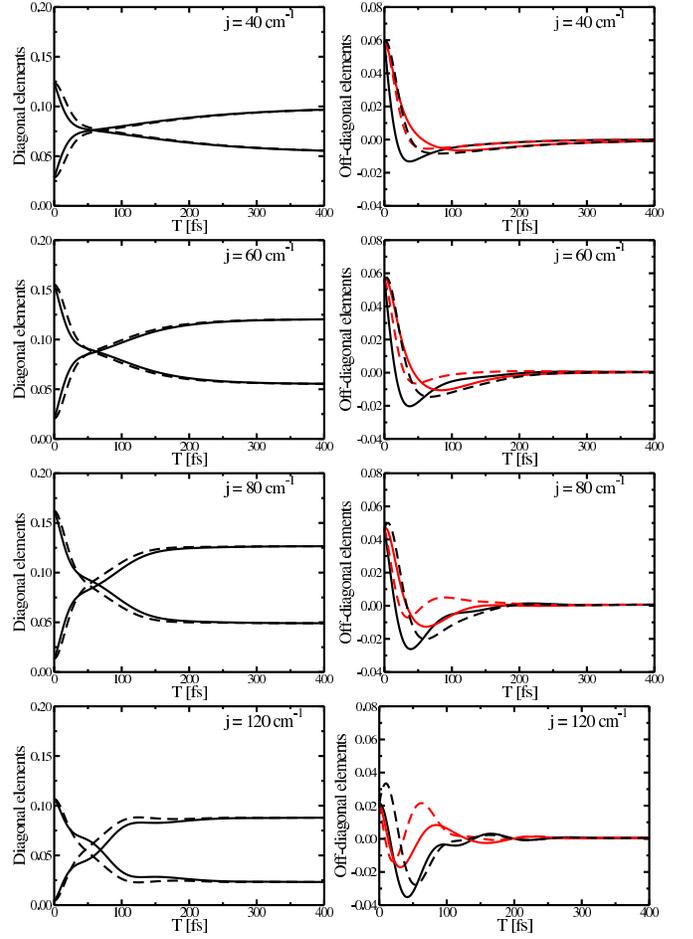} \caption{The same as Fig. \ref{fig:3} but for a heterodimer with parallel
transition dipole moment arrangement.\label{fig:4}}

\end{figure}

\begin{figure}[h]
\centering{}\includegraphics[clip,width=1\columnwidth]{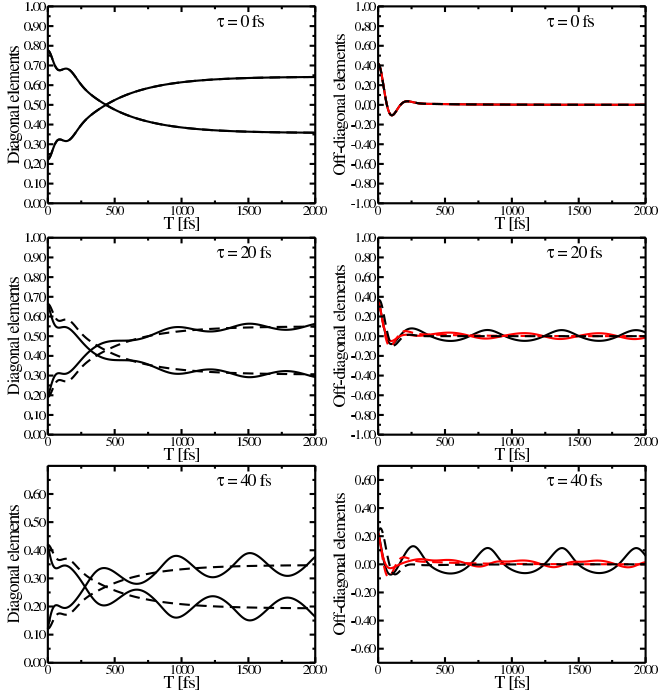} \caption{The time evolution of the real parts of $R_{I}$ and $R_{II}$ a heterodimer
with energy gap $\Delta=100\;\mathrm{cm}^{\mbox{-}1}$, resonance
coupling $j=33.1\;\mathrm{cm^{\mbox{-}1}}$ and a parallel arrangement
of the transition dipole moments, interacting with a bath represented
by one overdamped Brownian oscillator with parameters $\Lambda^{-1}=100\;\mathrm{fs}$
and $\lambda_{\mathrm{overdamped}}=12\;\mathrm{cm^{\mbox{-}1}}$ and
an underdamped vibration with reorganization energy $\lambda_{\mathrm{harmonic}}=30\;\mathrm{cm^{\mbox{-}1}}$
and frequency $\Omega_{\mathrm{Bath}}=60\;\mathrm{cm^{\mbox{-}1}}$.
Black lines, the matrix elements $11$, $22$ (left column) and $12$
(right column) of $R_{II}$ according to standard CL-QME (dashed)
and the corresponding parametric CL-QME (full). Red lines, the matrix
element $12$ (right column) of $R_{I}$ according to standard CL-QME
(dashed) and the corresponding parametric CL-QME (full).\label{fig:new1}}

\end{figure}

The most simple system which exhibits $\tau$-dependent correction
to the standard CL-QME is a molecular heterodimer. In general, it
is characterized by orientation and magnitude of its transition dipole
moments, the excited state energies $\varepsilon_{1}$, $\varepsilon_{2}$
of the component molecules, their resonance coupling $J$ and the
properties of the bath. We denote the difference of the excited state
energies by $\Delta\equiv\varepsilon_{1}-\varepsilon_{2}$, and we
set the magnitudes of the transition dipole moments to unity. The
initial condition is assumed in a form $\rho^{(1)}(0)=|g\rangle\langle g|$.
The evolution during the first interval of the response follows Eq.
(\ref{eq:I_interv}). For the calculations presented on Figs. \ref{fig:1},
\ref{fig:2} and \ref{fig:3}, we choose the anti-parallel orientation
of the transition dipole moments, while for the calculation shown
on Figs. \ref{fig:4} and \ref{fig:new1}, we choose the parallel
orientation. The temperature is set to $T=300$ K in all calculations.

First, let us investigate the sensitivity of the excited state dynamics
to the interplay of the delay $\tau$ and the phase of the bath vibrations.
Fig. \ref{fig:1} shows the dynamics of the electronic coherence between
the excited states $|e_{1}\rangle$ and $|e_{2}\rangle$ in presence
of the bath represented by a single-mode general Brownian oscillator
with parameters $\lambda=50\;\mathrm{cm^{\mbox{-}1}}$, $\gamma=1\;\mathrm{ps^{\mbox{-}1}}$,
$\Omega_{\mathrm{Bath}}=100\;\mathrm{ps^{\mbox{-}1}}$ and $\Delta=100\;\mathrm{cm}^{\mbox{-}1}$.
Both calculations show that the standard CL-QME calculation of ${\cal U}^{(eeee)}(T)$
is insensitive to the phase of bath vibration mode during the time
evolution in the first interval. The parametric CL-QME, Eq. (\ref{eq:Final-Int}),
shows a distinct sensitivity to this phase. In both cases, the resonance
coupling $J$ is chosen to be zero, and the dynamics can therefore
be evaluated exactly by the cumulant expansion technique (see Appendix
\ref{sec:Appendix}). The numerical evaluation using Eq. (\ref{eq:Final-Int})
indeed matches the analytical result. In the calculation on the Fig.
\ref{fig:1}, the vibration of the bath is much faster than the period
(333 fs) of the electronic coherence. The two theories give the same
result for $\tau=0\;\mathrm{fs}$, then they start to deviate and
after one period of bath oscillator, at $\tau=60\;\mathrm{fs}$, they
coincide again. Initial phase of $\rho_{I,12}$ and $\rho_{II,12}$
is in general different at $T=0\;\mathrm{fs}$ because of their different
time evolution in the first interval (they are not simply complex
conjugates of each other). In the Fig. \ref{fig:2}, we calculated
the same system, but we used bath with parameters $\lambda=5\;\mathrm{cm^{\mbox{-}1}}$,
$\gamma=1\;\mathrm{ps^{\mbox{-}1}}$, $\Omega_{\mathrm{Bath}}=19\;\mathrm{ps^{\mbox{-}1}}$.
The frequency is now resonant with the frequency of the electronic
coherence, which makes the effect more significant. The two theories
give the same result for $\tau=0\;\mathrm{fs}$, and at $\tau=360\;\mathrm{fs}$,
which is approximately one period of the bath vibration mode. Unlike
in Fig.\ref{fig:1}, the initial phase of $\rho_{I,12}$ and $\rho_{II,12}$
differs significantly in $T=0\;\mathrm{fs}$ because of their different
time evolution in $\tau$. 

In both the cases studied above, the time evolution of the off-diagonal
elements of the second order response operator is slightly modulated
by the time evolution of the vibrational DOF. The phase of the oscillations
seems to be mostly unaffected.

Figs. \ref{fig:3} and \ref{fig:4} demonstrate the influence of the
resonance coupling on the population and electronic coherence dynamics
in the homodimer. As above, we perform calculation according to standard
CL-QME and the parametric CL-QME. This time, we choose the overdamped
Brownian oscillator with fixed $\tau=60\;\mathrm{fs}$ and $\Delta=100\;\mathrm{cm}^{\mbox{-}1}$,
reorganization energy $\lambda=120\;\mathrm{cm^{\mbox{-}1}}$ and
correlation time $\tau_{c}=\Lambda^{-1}=50\;\mathrm{fs}$ to represent
the bath, and we change the resonance coupling. We calculate both
diagonal and off-diagonal elements ({}``populations'' and {}``coherences'')
of the operators $\rho_{I/II}$. For $J=0\;\mathrm{cm^{\mbox{-}1}}$,
there is no population dynamics. By increasing the coupling, the difference
between the theories in the diagonal elements increases. In the Fig.
\ref{fig:3}, the dipole moments of the molecules are anti-parallel,
while in Fig. \ref{fig:4} they are parallel. 


\begin{figure}[h]
\centering{}\includegraphics[clip,width=0.7\columnwidth]{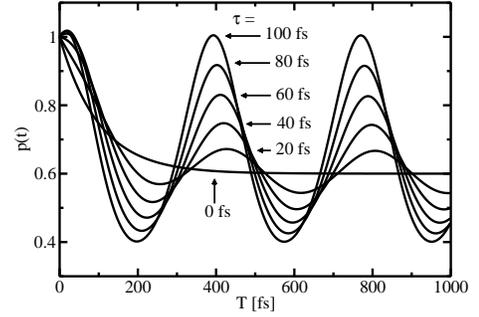} \caption{The time evolution of {}``population'' according to model Eq. (\ref{eq:simple-model-equation}).
The curves are solutions for $\tau$ increasing from 0 fs to 100 fs
with step of 20 fs. \label{fig:new2}}

\end{figure}

Let us now investigate a molecular dimer coupled to overdamped harmonic
bath and to a single harmonic mode with frequency $\Omega_{\mathrm{Bath}}$.
The harmonic mode is assumed to continue oscillating even long after
thermalization in the overdamped part of the bath has taken place.
Therefore, the $\tau$-dependent term of Eq. (\ref{eq:New-terms-final})
changes the system dynamics also at long times. The time dependence
of the density operator elements for different $\tau$ is shown in
Fig. \ref{fig:new1}. Parameters of the overdamped bath are $\Lambda^{-1}=100\;\mathrm{fs}$
and $\lambda_{\mathrm{overdamped}}=12\;\mathrm{cm^{\mbox{-}1}}$ and
of the harmonic mode $\lambda_{\mathrm{harmonic}}=30\;\mathrm{cm^{\mbox{-}1}}$,
$\Omega_{\mathrm{Bath}}=60\;\mathrm{cm^{\mbox{-}1}}$. The dimer is
characterized by $\Delta=100\;\mathrm{cm^{\mbox{-}1}}$, $J=33.1\;\mathrm{cm^{\mbox{-}1}}$
and the parallel electronic transition dipole moments of the molecules.
We can notice that the interaction of the vibrational mode with the
electronic DOF induces oscillations in both the diagonal and off-diagonal
elements of the density operator. The amplitude of the oscillations
increases with increasing $\tau$. Since the EGCF of the harmonic
mode, Eq. (\ref{eq:Cosc}), is periodic, we expect the relative amplitude
of the oscillations to decrease again for sufficiently long $\tau$,
and to become zero for $\tau=2\pi/\Omega_{\mathrm{Bath}}$. However,
in such long $\tau$ the second order response would be almost zero
due to the decay of the optical coherence in the first interval. Fig.
\ref{fig:new1} suggests that the phase of the oscillation does not
change linearly with $\tau$. To understand this behavior, we can
study a simple equation of the form 

\begin{equation}
\dot{\rho}(t)=-\alpha(\rho(t)-\rho_{0})+(f(t+\tau)-f(t))\rho(t),\label{eq:simple-model-equation}
\end{equation}
which describes exponential decay to a limiting value $\rho_{0}$
with the rate constant $\alpha$ and a modulation by the function
\begin{equation}
f(t)=\int\limits _{0}^{t}d\tau\; A\cos(\omega\tau).\label{eq:ft}
\end{equation}
Eq. (\ref{eq:ft}) represents a first integral of the EGCF of an underdamped
vibrational mode, Eq. (\ref{eq:Cosc}). The solution of Eq. (\ref{eq:ft})
with the parameters $\alpha=0.01$, $A=10^{-4}$, $\omega^{-1}=60$
and $\rho_{0}=0.6$ is shown in Fig. \ref{fig:new2}. We can see that
the oscillation phase and period of the oscillations is indeed not
proportional to $\tau$, and the amplitude increases with $\tau$
similarly to Fig. \ref{fig:new1}. The behavior is therefore a direct
consequence of the parametric term in the parametric QME.

The overall picture arising from the numerical simulations is the
following: Except for special cases, such as the molecular homodimer,
the excited state dynamics of an open quantum system, as it is observed
by the non-linear spectroscopy, indeed depends on the delay $\tau$
between the FWM scheme. The 2D spectroscopy therefore observes a certain
averaged dynamics. The effects seem to be rather small in most cases,
but they might be observable by an advanced implementation of 2D spectroscopy.
They are especially pronounced in the case of the intramolecular vibrational
modes, which have frequency similar to the electronic energy gap between
excitonic levels. Both the dynamics of electronic level populations
and electronic coherences are affected. In order to identify these
effects in the experimental data, the theory has to be extended to
include also the third interval of the third order non-linear response.
The corresponding projection operator ${\cal P}_{T+\tau}$ which now
depends on the duration of both the coherence and the population intervals
$\tau$ and $T$, respectively, has been already proposed an tested
for $j_{mn}=0$ in Ref. \cite{Mancal2010c}. The formulation of the
theory for the third interval of the response of a multilevel excitonic
system in a similar manner as performed in this paper for the second
interval will be the subject of our future work. 


\section{Conclusions}

In this paper, we have demonstrated that the third and the second
order non-linear response of a multilevel system cannot be completely
evaluated by propagating reduced density matrix by equations of motion
derived using a single projection operator. Such treatment would inevitably
neglect correlations between the time evolution of the bath during
the neighboring intervals of the non-linear response. We have derived
equation of motion, the parametric quantum master equation, which
takes these correlations into account approximately, and and showed
that in the absence of resonance coupling the method yields an agreement
with the result obtained by the second order cumulant method. We confirm
by numerical simulations that for different delays $\tau$ between
the excitation pulses, distinct dynamics of both excite state populations
and electronic coherence occurs, in the presence of environmental
degrees of freedom with finite bath correlation time and in the presence
of intramolecular vibrations. The two-dimensional Fourier transformed
spectroscopy sees in these cases some averaged dynamics. 
\begin{acknowledgments}
This work was supported by the Czech Science Foundation (GACR) grant
nr. 205/10/0989 and the Ministry of Education, Youth and Sports of
the Czech Republic via grant KONTAKT ME899 and the research plan MSM0021620835.
J. O. acknowledges the support by grant GAUK 102-10/251386. 
\end{acknowledgments}

\appendix

\section{Third-Order Non-linear Response Functions\label{sec:Third-Order-Non-linear-Response}}

The four standard response functions (so-called Liouville pathways)
of a two-band electronic system read in our block formalism as

\[
R_{1}(t,T,\tau)={\rm tr}\{\mu^{(ge)}{\cal U}^{(egeg)}(t){\cal V}^{(egee)}
\]
\begin{equation}
\times{\cal U}^{(eeee)}(T){\cal V}^{(eeeg)}{\cal U}^{(egeg)}(\tau){\cal V}^{(eggg)}W_{\mathrm{eq}}^{(gg)}\},\label{eq:R1-1}
\end{equation}
\[
R_{2}(t,T,\tau)={\rm tr}\{\mu^{(ge)}{\cal U}^{(egeg)}(t){\cal V}^{(egee)}
\]
\begin{equation}
\times{\cal U}^{(eeee)}(T){\cal V}^{(eege)}{\cal U}^{(gege)}(\tau){\cal V}^{(gegg)}W_{\mathrm{eq}}^{(gg)}\},\label{eq:R2-1}
\end{equation}
\[
R_{3}(t,T,\tau)={\rm tr}\{\mu^{(ge)}{\cal U}^{(egeg)}(t){\cal V}^{(eggg)}
\]
\begin{equation}
\times{\cal U}^{(gggg)}(T){\cal V}^{(ggge)}{\cal U}^{(gege)}(\tau){\cal V}^{(gegg)}W_{{\rm \mathrm{eq}}}^{(gg)}\},\label{eq:R3-1}
\end{equation}
\[
R_{4}(t,T,\tau)={\rm tr}\{\mu^{(ge)}{\cal U}^{(egeg)}(t){\cal V}^{(eggg)}
\]
\begin{equation}
\times{\cal U}^{(gggg)}(T){\cal V}^{(ggeg)}{\cal U}^{(egeg)}(\tau){\cal V}^{(eggg)}W_{\mathrm{eq}}^{(gg)}\}.\label{eq:R4-1}
\end{equation}
Here, $W_{{\rm eq}}^{(gg)}=w_{{\rm eq}}|g\rangle\langle g|$ where
$w_{{\rm eq}}$ is the bath equilibrium density operator.

\section{Derivation of Relaxation Terms\label{sec:Derivation-of-Relaxation}}

In this section we will evaluate the term ${\rm NZ}_{3}$ of Eq. (\ref{eq:nakajima-zwanzig}).
Applying the definitions of its component operators from Section \ref{sec:Reduced-density-matrix}
we obtain

\begin{equation}
\mathrm{NZ_{3}}=-\frac{i}{\hbar}\mathrm{tr}_{B}\Bigg\{\left[\sum_{m}\Delta\Phi_{m}(t)K_{m}(t),\rho^{(I)}(t)\, w_{\tau}^{(I)}\right]_{-}\Bigg\} w_{\tau}^{(I)}
\end{equation}
where 
\begin{equation}
\rho^{(I)}(t)=\mathrm{tr}_{B}\left\{ \mathcal{P}_{\tau}W^{(I)}(t)\right\} .
\end{equation}
Using Eq. (\ref{eq:wItau_def}) yields

\begin{align}
\mathrm{NZ_{3}}= & -\frac{i}{\hbar}\sum_{m\tilde{k}}e^{g_{\tilde{k}}^{*}(\tau)}\mathrm{tr}_{B}\Big\{\Delta\Phi_{m}(t)K_{m}(t)\rho^{(I)}(t)\, w_{\mathrm{eq}}\nonumber \\
 & \;\;\;\times U^{g}(\tau)U_{\tilde{k}}^{e\dagger}(\tau)K_{\tilde{k}}\Big\} w_{\tau}^{(I)}\nonumber \\
\, & +\frac{i}{\hbar}\sum_{m\tilde{k}}e^{g_{\tilde{k}}^{*}(\tau)}\mathrm{tr}_{B}\Big\{\rho^{(I)}(t)\, w_{\mathrm{eq.}}U^{g}(\tau)U_{\tilde{k}}^{e\dagger}(\tau)\nonumber \\
 & \;\;\;\times K_{\tilde{k}}\Delta\Phi_{m}(t)K_{m}(t)\Big\} w_{\tau}^{(I)}\;.
\end{align}
Now we have to set $e^{g_{\tilde{k}}^{*}(\tau)}\approx1$ since its
contribution is of higher order in $\Delta\Phi$. Applying the first
order expansion 
\begin{equation}
U^{g}(\tau)U_{\tilde{n}}^{e\dagger}(\tau)\approx1+\frac{i}{\hbar}\int\limits _{0}^{\tau}d\tau'\,\Delta\Phi_{\tilde{n}\tilde{n}}(-\tau'),\label{eq:UgUe_expansion}
\end{equation}
yields

\begin{align}
\mathrm{NZ_{3}}= & -\frac{i}{\hbar}\sum_{m\tilde{k}}\mathrm{tr}_{B}\Bigg\{\Delta\Phi_{m}(t)K_{m}(t)\rho^{(I)}(t)\, w_{\mathrm{eq}}\nonumber \\
 & \;\;\;\times\Bigg(1+\frac{i}{\hbar}\int\limits _{0}^{\tau}d\tau'\Delta\Phi_{\tilde{k}\tilde{k}}(-\tau')\Bigg)K_{\tilde{k}}\Bigg\} w_{\tau}^{(I)}\nonumber \\
\, & +\frac{i}{\hbar}\sum_{m\tilde{k}}\mathrm{tr}_{B}\Bigg\{\Bigg(1+\frac{i}{\hbar}\int\limits _{0}^{\tau}d\tau'\Delta\Phi_{\tilde{k}\tilde{k}}(-\tau')\Bigg)\nonumber \\
 & \;\;\;\times K_{\tilde{k}}\Delta\Phi_{m}(t)K_{m}(t)\rho^{(I)}(t)\, w_{\mathrm{eq}}\Bigg\} w_{\tau}^{(I)}.
\end{align}
Now, we define a line shape function 
\begin{equation}
g_{a\tilde{b}}(t)=\frac{1}{\hbar^{2}}\int\limits _{0}^{t}d\tau\int\limits _{0}^{\tau}\tau'\;\langle\Delta\Phi_{a}(\tau')\Delta\Phi_{\tilde{b}}\rangle,\label{eq:g-function-alternative}
\end{equation}
where it is noteworthy that one of its indices represents the site-
and the other the exciton-basis. The ${\rm NZ}_{3}$ term can then
be expressed as

\begin{align}
\mathrm{NZ_{3}}=\; & \frac{1}{\hbar^{2}}\int\limits _{0}^{\tau}d\tau'\Bigg[\sum_{m\tilde{k}}K_{m}(t)\rho^{(I)}(t)K_{\tilde{k}}-\rho^{(I)}(t)K_{\tilde{k}}K_{m}(t)\Bigg]\nonumber \\
 & \;\;\;\times\mathrm{tr}_{B}\left\{ \Delta\Phi_{\tilde{k}}(-\tau')\Delta\Phi_{m}(t)\, w_{\mathrm{eq}}\right\} w_{\tau}^{(I)}\nonumber \\
= & \sum_{m\tilde{k}}\,\left(K_{m}(t)\rho^{(I)}(t)K_{\tilde{k}}-\rho^{(I)}(t)K_{\tilde{k}}K_{m}(t)\right)\nonumber \\
 & \;\;\;\times\left(\dot{g}_{m\tilde{k}}^{*}(t+\tau)-\dot{g}_{m\tilde{k}}^{*}(t)\right)w_{\tau}^{(I)}.\label{eq:NZ3-new-term-derived}
\end{align}
In order to obtain the last two lines of Eq. (\ref{eq:Final-Int})
we will trace Eq. (\ref{eq:NZ3-new-term-derived}) over the bath DOF.

\section{The $\tau$-dependent Term for a Homodimer\label{sec:Homodimer}}

Let us assume a molecular homodimer with the Hamiltonian
\begin{align}
H_{S} & =\left[\begin{array}{cc}
0 & j\\
j & 0
\end{array}\right]\;,\\
H_{S-B} & =\left[\begin{array}{cc}
\Delta\Phi_{1} & 0\\
0 & \Delta\Phi_{2}
\end{array}\right].
\end{align}
This yields in the exciton basis
\begin{align}
H_{S} & =\left[\begin{array}{cc}
-j & 0\\
0 & j
\end{array}\right]\;,\\
H_{S-B} & =\frac{1}{2}\left[\begin{array}{cc}
\Delta\Phi_{1}+\Delta\Phi_{2} & \Delta\Phi_{2}-\Delta\Phi_{1}\\
\Delta\Phi_{2}-\Delta\Phi_{1} & \Delta\Phi_{1}+\Delta\Phi_{2}
\end{array}\right].
\end{align}
The derivative of the line shape function which enters the $\tau$-dependent
correction ${\cal D}(t;\tau)$ of the CL-QME then reads as 
\begin{equation}
\dot{g}_{m\tilde{k}}(t)=\frac{1}{2}\int\limits _{0}^{t}{\rm d}t^{\prime}\sum_{n=1}^{2}\langle\Delta\Phi_{m}(t^{\prime})\Delta\Phi_{n}\rangle.
\end{equation}
We neglect the cross-terms $\langle\Delta\Phi_{i}(t)\Delta\Phi_{j}\rangle$
for $i\neq j$ , because we assume only local correlations of the
bath. In a homodimer $\Delta\Phi_{1}=\Delta\Phi_{2}$, and we get
$\dot{g}_{1\tilde{k}}(t)=\dot{g}_{2\tilde{k}}(t)=\dot{g}(t)$ from
which it follows that
\begin{equation}
\dot{g}_{m\tilde{k}}^{*}(t+\tau)-\dot{g}_{m\tilde{k}}^{*}(t)=\dot{g}^{*}(t+\tau)-\dot{g}^{*}(t).
\end{equation}
By substituting this equation into Eq. (\ref{eq:New-terms-final}),
the $\tau-$containing term ${\cal D}(t;\tau)$ turns to zero, because
\begin{equation}
\sum_{m\tilde{k}}\,\left(K_{m}\rho(t)K_{\tilde{k}}-\rho(t)K_{\tilde{k}}K_{m}\right)=0.
\end{equation}
Thus the $\tau$-dependent correction to the CL-QME is zero for the
case of a homodimer.

\section{Pure Dephasing of an Electronic Coherence\label{sec:Appendix}}

In a pure dephasing model of the system-bath interaction, analytical
solution for the dynamics of electronic coherences can be obtained
with the cumulant expansion technique. The time evolution of coherence
$\rho_{ee'}^{(2)}(t;\tau)$ reads as 
\begin{equation}
\rho_{ee'}^{(2)}(t,\tau)=\mathrm{tr}_{B}\left\{ U_{e}(t)U_{g}(\tau)W_{\mathrm{eq}}U_{e'}^{\dagger}(\tau)U_{e'}^{\dagger}(t)\right\} \rho_{ee'}^{(2)}(0),
\end{equation}
where we set all transition dipole moments to one. By using the second
cumulant procedure \cite{MukamelBook} we can evaluate the expression
into 
\begin{align}
\rho_{ee'}^{(2)}(t;\tau)= & \; e^{-g_{ee}(t)-g_{e'e'}^{*}(t+\tau)+g_{e'e}(t)}\nonumber \\
 & e^{+g_{ee'}^{*}(t+\tau)-g_{ee'}^{*}(\tau)}\rho_{ee'}^{(2)}(0).\label{eq:cumulant-rho_ee_exp}
\end{align}
Equivalently, we can rewrite Eq. (\ref{eq:cumulant-rho_ee_exp}) in
form of a master equation 

\begin{align}
\frac{d}{dt}\rho_{ee'}^{(2)}(t;\tau)= & \;\Big[-\dot{g}_{e'e'}^{*}(t)-\dot{g}_{ee}(t)\nonumber \\
 & +\dot{g}_{e'e}(t)+\dot{g}_{ee'}^{*}(t)\nonumber \\
\, & -\left(\dot{g}_{e'e'}^{*}(t+\tau)-\dot{g}_{e'e'}^{*}(t)\right)\nonumber \\
 & +\left(\dot{g}_{ee'}^{*}(t+\tau)-\dot{g}_{ee'}^{*}(t)\right)\Big]\rho_{ee'}^{(2)}(t;\tau).\label{eq:ex_meq}
\end{align}
The first four terms on the right hand side of Eq. (\ref{eq:ex_meq})
are described by the CL-QME. The remaining terms can be matched with
the term ${\cal D}(t;\tau)$ derived in this paper. Eq. (\ref{eq:Final-Int})
therefore gives the correct result for this simple exactly solvable
case. 

\bibliographystyle{prsty}

\end{document}